\documentstyle[12pt]{article}
\title{A formula with volumes of five tetrahedra and discrete curvature}
\author{Igor G. Korepanov\thanks{South Ural State University,
76 Lenin ave., Chelyabinsk 454080, Russia.
E-mail address: igor@prima.tu-chel.ac.ru}}
\date{}

\def\be{\begin{equation}}
\def\ee{\end{equation}}
\def\ol{\overline}
\def\pa#1#2{{\partial#1\over\partial#2}}

\begin{document}
\maketitle

\begin{abstract}
Given five points in a three-dimensional euclidean space, one can consider
five tetrahedra, using those points as vertices. We present a
pentagon-like formula containing the product of three volumes of those
tetrahedra in its l.h.s.\ and the product of the two remaining tetrahedron
volumes in its r.h.s., as well as the derivative of the ``discrete
curvature'' which arises when we slightly deform our euclidean space.
\end{abstract}

In this short note, we derive an easy consequence from the duality
formulae presented in~\cite{3s-t}, namely, some formula that can
be treated as a sort of pentagon equation involving five tetrahedra
in a three-dimensional space. Hopefully, such formulas may be of use
for both geometry (manifold invariants) and statistical physics (given
that {\em positive\/} Boltzmann weights arise almost ``automatically''),
especially because of the possibility of their wide generalization,
including ``spherical'' case and, very probably, higher dimensions.

Consider five points $A$, $B$, $C$, $D$ and $E$ in the three-dimensional
euclidean space. There exist ten distances between them, which we will
denote as $l_{AB}$, $l_{AC}$ and so on.

Let us fix all the distances except $l_{AB}$ and  $l_{DE}$. Then,
$l_{AB}$ and  $l_{DE}$ satisfy one constraint (Cayley -- Menger equation)
which we can, using arguments like those in~\cite{3s-t}, represent
in the following differential form:
\be
\left| l_{AB} \,dl_{AB} \over V_{\ol D}\, V_{\ol E} \right| =
\left| l_{DE} \,dl_{DE} \over V_{\ol A}\, V_{\ol B} \right|,
\label{eq AB-DE}
\ee
where, say, $V_{\ol A}$ denotes the volume of tetrahedron $\ol A$, that is
one with vertices $B$, $C$, $D$ and $E$ (and {\em without\/}~$A$).

Let us consider the dihedral angles at the edge $DE$ --- the common edge
for tetrahedra $\ol A$, $\ol B$ and $\ol C$. Namely, denote
$$
\angle BDEC \stackrel{\rm def}{=} \alpha, \quad
\angle CDEA \stackrel{\rm def}{=} \beta, \quad
\angle ADEB \stackrel{\rm def}{=} \gamma.
$$
We have:
\be
0=d(\alpha+\beta+\gamma) = \pa{\gamma}{l_{AB}} \,dl_{AB} +
\pa{(\alpha+\beta+\gamma)}{l_{DE}} \,dl_{DE}.
\ee

According to~\cite[formula~(11)]{3s-t},
\be
\left|\pa{\gamma}{l_{AB}}\right|={1\over 6}\left| l_{AB}\,l_{DE}\over
V_{\ol C}\right|.
\ee
Denote also
\be
\alpha+\beta+\gamma \stackrel{\rm def}{=} 2\pi-\omega,
\label{eq dgamma}
\ee
where $\omega$ is the ``defect angle''. The formulas
(\ref{eq AB-DE}--\ref{eq dgamma}) together yield
\be
{1\over 6}\, \left| l_{DE}^2\over V_{\ol A}\, V_{\ol B}\, V_{\ol C} \right|
= \left|{1\over V_{\ol D}\, V_{\ol E}}\, \pa{\omega}{l_{DE}} \right|,
\ee
which can be also written by means of the following integral in the
length of the edge~$DE$ ``redundant'' for the tetrahedra $\ol D$ and
$\ol E$:
\be
{1\over 6}\, \left| \int {\delta(\omega)\, l_{DE}^2 \,dl_{DE}
\over V_{\ol A}\, V_{\ol B}\, V_{\ol C} } \right| =
\left|{1\over V_{\ol D}\, V_{\ol E}} \right|,
\ee
with the integral taken over a neighborhood of the value of~$l_{DE}$
corresponding to the flat space ($\omega=0$); $\delta$ is the Dirac
delta function.

\subsubsection*{Remarks}

\noindent {\bf 1.}\quad
Similar considerations will be much more complicated already for the
four-dimensional space, where we will have 6 vertices, 6 tetrahedra,
15 edges, and 20 two-dimensional faces where the ``discrete curvature''
can be concentrated.

\medskip

\noindent {\bf 2.}\quad
It may seem that the whole manifold where everything happens must be
``flat'' due to the delta function $\delta(\omega)$. However, it is not so,
and not only because of the mentioned possibility of generalization to
the spherical geometry. Imagine, for instance, a ramified covering
of some flat manifold, where the ``full angle'' corresponding to going
around a small contour surrounding a
``ramification contour'' can equal any multiple of~$2\pi$.

\end{document}